\documentclass[a4paper,11pt]{article}
\usepackage{theorem}
\theorembodyfont{\itshape}
\newtheorem{pro}{Proposition}
\theorembodyfont{\rmfamily}
\newtheorem{pf}{Proof}

\usepackage{amssymb}
\usepackage{latexsym}
\usepackage{amsmath}
\usepackage{braket}
\usepackage[dvips]{graphicx}
\usepackage{cite}
\title{Existence of equilibria in quantum Bertrand--Edgeworth\\ duopoly game}
\author{Yohei Sekiguchi,$^1$ Kiri Sakahara,$^1$ Takashi Sato $^{1,2}$\\
$^1$ {\normalsize\textit{Graduate School of Economics, University of Tokyo, 7-3-1 Hongo, Bunkyo-ku, Tokyo, Japan}}\\
$^2$ {\normalsize\textit{Faculty of Economics, Toyo University, 5-28-20 Hakusan, Bunkyo-ku, Tokyo, Japan}}
}
\date{March 2010}
\begin{document}
\maketitle

\begin{abstract}
Both classical and quantum version of two models of price competition in duopoly market, the one is realistic and the other is idealized, are investigated.
The pure strategy Nash equilibria of the realistic model exists under stricter condition than that of the idealized one in the classical form game.
This is the problem known as Edgeworth paradox in economics. In the quantum form game, however, the former converges to the latter as the measure of entanglement goes to infinity.

PACS numbers: 03.67.-a, 02.50.Le
\end{abstract}

\section{Introduction}
Game theory is a powerful mathematical tool to analyze various natural and social phenomena \cite{Gibbons, Axelrod, Maynard_Smith}.
After the publication of Meyer \cite{M99}, there has been a great deal of effort to extend the classical game theory into the quantum domain, and it has been shown that quantum games may have significant advantages over their classical counterparts \cite{M99, EWL99, BH01}.
\textit{Nash equilibrium} \cite{Nash}, a point where players cannot be better off by unilaterally changing their actions, is the most important solution concept in both the classical and the quantum game theory.
However, existence of Nash equilibria is not always ensured, e.g., the penny flip game or the rock--scissors--paper game.
Meyer \cite{M99} considered a quantum version of the penny flip game, and showed that quantum extensions may resolve nonexistence of Nash equilibria in classical games (see also \cite{IT02, IT04, F04, I05}).
We also attack this problem by analyzing a quantum extension of a game which describes market competition.

In economics, many important markets are neither perfectly competitive nor perfectly monopolistic, that is, the action of individual firms affect the market price \cite{Tirole}.
These markets are usually called \textit{oligopolistic} and can be analyzed based on game theory.
Two of the oldest and most important models describing competition among firms in a oligopolistic market are the \textit{Cournot model} \cite{Cournot} and the \textit{Bertrand model} \cite{Bertrand}.
The difference between these models is strategic variables chosen by competing firms:
in the former, firms select \textit{quantities} of their products, whereas in the latter they choose \textit{prices} at which they are willing to sell their products.
It is worth pointing out that each firm cannot set both price and quantity in oligopolistic markets since these are related to each other by the demand curve, yet changing the strategic variables significantly alters the results obtained.
Under the assumption that each firm can produce a homogeneous product with the constant marginal cost as much as they want, the Cournot model predicts that the equilibrium price lies  between the competitive equilibrium (hereafter CE) price and the monopoly price.
On the other hand, the Bertrand model asserts that the equilibrium price must be equal to the CE price, at which firms earn zero profit, under the same assumption.
In the literature this result is known as the \textit{Bertrand paradox}, since it suggests that competition among only two firms may be sufficient to yield the perfectly competitive outcome.

Efforts at resolving the Bertrand paradox have involved relaxing the various assumptions underlying the model, e.g. that the cost functions are linear, or that the product is homogeneous, etc.
Edgeworth \cite{Edgeworth} is the earliest attack to resolve the Bertrand paradox by introducing \textit{capacity constraint} by which firms cannot sell more than they are capable of producing.
Of course, if the capacity is sufficiently large so that the constraint is not binding, then the results are unchanged from the original Bertrand model.
On the other hand, as shown by Edgeworth \cite{Edgeworth}, if the capacity is sufficiently small, then there exists the unique equilibrium, where the equilibrium price equals the CE price and firms obtain positive profit.
However, if the capacity lies in the intermediate range, then it causes another problem called the \textit{Edgeworth paradox}: pure strategy equilibria fail to exist.\footnote{It is commonly known that there exist mixed strategy equilibria, even if pure ones fail to exist \cite{DM86}.}
Hence, in such an industry, the market price will fluctuate and the social surplus will not be maximized.

To complete the model explaining the Bertrand--Edgeworth competition, we need to specify the \textit{residual demand functions} which determine firms' residual demand given prices charged by the firms.
In the literature, two of the most commonly used rules to determine the residual demands are the \textit{efficient rationing} rule and the \textit{proportional rationing} rule \cite{Tirole}.
As explained below, the proportional rationing is said to be more suitable to describe real markets than the efficient rationing: the efficient rationing portrays an ideal market where the social surplus is always maximized.
However, the existence of equilibria is, unfortunately, more severe under the proportional rationing than under the efficient rationing, that is, in real markets equilibria are hard to exist than in the idealized one.
Our question is whether or not the quantum entanglement helps the existence of equilibria in real markets.

Recently, Li et al. \cite{LDM02} investigated the quantization of games with continuous strategic space, a classic instance of which is the Cournot duopoly.
They showed that the firms can escape the frustrating dilemma--like situation if the structure involves a maximally entangled state.
In this paper, we apply Li et al.'s \cite{LDM02} ``minimal'' quantization rules to the Bertrand--Edgeworth duopoly, in which two identical capacity constrained firms compete in price.\footnote{Several quantum extensions of oligopolistic competition, applying Li et al.'s \cite{LDM02} ``minimal'' quantization rules, have been considered, e.g., the quantum Cournot duopoly game \cite{DLJ03, DJL05, SSS10}, the quantum Bertrand duopoly game with product differentiation \cite{LK04, QCSD05}, the quantum Stackelberg duopoly game \cite{LK03a, LK05}, and the quantum oligopoly game \cite{LK03b}.}
To consider a realistic market, we adopt the proportional rationing rule.
We observe the transition of the game from purely classical to fully quantum as the game's entanglement increases from zero to maximum.
We show that the range of capacities in which pure strategy equilibria fail to exist contracts as the entanglement increases.
In the maximally entangled game, the pure strategy equilibrium is uniquely determined, where both firms charge the CE price whenever it exists in the ideal market.
In other words, nonexistence in the classical Bertrand--Edgeworth duopoly is resolved and the social surplus is maximized in our quantum extension.

\section{Classical Bertrand--Edgeworth Duopoly}
There are two identical firms, Firm 1 and Firm 2, producing a homogeneous good.
Each firm can produce the good with zero cost up to the capacity constraint $k>0$.\footnote{All results of this analysis can be easily generalized to the case of constant marginal cost by considering the price in the analysis as the unit margin over cost.}
The market demand is given by
\begin{equation}
	D(p)=a-p.
\end{equation}
We assume that $k<a/2$ so that the CE price $\hat{p}=a-2k>0$, which implies the firms can obtain positive profit even in the competitive equilibrium.
The firms simultaneously charges prices $p_1$ and $p_2$.
It is assumed that the firm quoting the lower price serves the entire market up to its capacity, and that the residual demand is met by the other firm.
The residual demand depends, of course, on which consumers purchase from which firm.
We assume that each firm's residual demand $R_i(p_1,p_2)$ is determined by the \textit{proportional rationing} rule\cite{Tirole, Edgeworth, DM86}, that is,
\begin{eqnarray}
	R_i(p_1,p_2)=
		\begin{cases}
			D(p_i) & p_i< p_j \\
			D(p_i)/2 & p_i = p_j \\
			\max\{0, D(p_i)[1-k/D(p_j)]\} & p_i> p_j.
		\end{cases}
\end{eqnarray}
The proportional rationing rule describes the following situation.
Consumers take the firms' choices as given and try to purchase the good as cheaply as possible.
If the firm charging lower price has insufficient supply, then consumers are rationed at this price.
The rationing at the lower price is on a first--come--first--serve basis, and we assume that all consumers have the same probability of being rationed.
Namely, for $p_i>p_j$ the probability of not being able to buy from firm $j$ is $(1-k/D(p_j))$, which represents $i$'s market share.
On the other hand, if the two firms set the same price, then they equally share the market demand.
Firms can sell the good up to their residual demand, so long as their capacities are not met.
Thus, each firm's profit is given by
\begin{equation}
	u_i(p_1,p_2)=p_i\cdot \min\{k,R_i(p_1,p_2)\}.
\end{equation}
It is well known that in the unique pure strategy Nash equilibrium, if exists, both firms charge the CE price $\hat{p}$ (see the Appendix).
However, there exists a certain range of capacities where the pure strategy equilibrium fails to exist.
In fact, the unique equilibrium exists under the proportional rationing if and only if $k\le a/4$.

Suppose that both firms charge the CE price $\hat{p}$.
We check that each firm cannot be better off by raising her price if and only if $k\le a/4$.\footnote{Note that each firm cannot be better off by undercuts, since each firm has already produced up to her full capacity at the price and so  cannot increase her sales by undercuts.}
Let $\tilde{u}_i(p_i)$ be firm $i$'s profit when $i$ charges $p_i\in [\hat{p},a)$ and the opponent charges $\hat{p}$
\begin{equation}
	\tilde{u}_i(p_i)=p_i(a-p_i)/2.
\end{equation}
Thus, we have
\begin{equation}
	\tilde{u_i}'(\hat{p})=\frac{1}{2}(4k-a),
\end{equation}
which is nonpositive if and only if $k\le a/4$.
Since $\tilde{u}_i$ is concave, the above assertion is valid.

As a benchmark, consider the case where the residual demand function is determined by the efficient rationing rule.
In this case, $R_i(p_1,p_2)$ is given by
\begin{equation}
	R_i(p_1,p_2)=
		\begin{cases}
			D(p_i) & p_i< p_j \\
			D(p_i)/2 & p_i = p_j \\
			\max\{0, D(p_i)-k\} & p_i> p_j.
		\end{cases}
\end{equation}
This can be obtained if the consumers could costlessly resell the good to each other.
This rule is called ``efficient'' because of the fact that a market where we can trade with zero transaction cost, that is, a perfectly competitive exchange market, is efficient.
Note that, however, a costless arbitrage between consumers is an extremely strong assumption, as in \cite{Tirole},
\begin{quote}
the existence of a frictionless arbitrage between consumers is a strong assumption.
(Recall that on the other side of the market, firms are assumed not to be able to change prices; thus, we may be putting too much friction on the supply side and too little on the demand side.) (p.213)
\end{quote} 
Therefore, the efficient rationing is said to be an idealistic artifact such as frictionless air.
In contrast, in a market described by the proportional rationing the consumers are prohibited, or not willing, to resell the good.
Thus, we can say that the proportional rationing rule describes more realistic situation.
In real market, of course, there are several ways to resell the good with transaction cost, e.g., internet auction, but this is beyond the scope of the present study.

Under the efficient rationing, if $i$ charges $p_i\in [\hat{p},a)$ and the opponent charges $\hat{p}$, then $i$'s profit is
\begin{equation}
\tilde{u}_i(p_i)=p_i(a-p_i-k),
\end{equation}
and we have
\begin{equation}
\tilde{u}_i'(p_i)=a-k-2p_i,
\end{equation}
which is nonpositive if and only if $k\le a/3$. 
Thus, under the efficient rationing, the unique equilibrium exists if and only if $k\le a/3$.
(The uniqueness result is similarly valid.)
Consequently, we can say that in real markets the equilibrium is hard to exist than in the idealized one.

\section{Quantum Bertrand--Edgeworth Duopoly}
To model the Bertrand--Edgeworth duopoly on a quantum domain, we follow Li et al.'s ``minimal'' extension, which utilizes two single--mode electromagnetic fields, of which the quadrature amplitudes have a continuous set of eigenstates.
The tensor product of two single--mode vacuum states $\ket{vac}_1\otimes \ket{vac}_2$ is identified as the starting state of the Cournot game, and the state consequently undergoes a unitary entanglement operation $\hat{J}(\gamma)\equiv \exp\{-\gamma(\hat{a}_1^\dagger\hat{a}_2^\dagger-\hat{a}_1\hat{a}_2)\}$, in which $\hat{a}_1$ and $\hat{a}_2$ ($\hat{a}_1^\dagger$ and $\hat{a}_2^\dagger$) are the annihilation (creation) operators of the electromagnetic field modes.
The operation is assumed to be known to both firms and to be symmetric with respect to the interchange of the two field modes.
The resultant state is given by $\ket{\psi_i}\equiv \hat{J}(\gamma)\ket{vac}_1\otimes\ket{vac}_2$.
Then firm $1$ and firm $2$ execute their strategic moves via the unitary operations $\hat{D}_1(x_1)\equiv \exp\{x_1(\hat{a}_1^\dagger-\hat{a}_1)/\sqrt{2}\}$ and $\hat{D}_2(x_2)\equiv \exp\{x_2(\hat{a}_2^\dagger-\hat{a}_2)/\sqrt{2}\}$, respectively, which correspond to the quantum version of the strategies of the Cournot game.
The final measurement is made, after these moves are finished and a disentanglement operation $\hat{J}(\gamma)^\dagger$ is carried out.
The final state prior to the measurement, thus, is $\ket{\psi_f}\equiv \hat{J}(\gamma)^\dagger\hat{D}_1(x_1)\hat{D}_2(x_2)\hat{J}(\gamma)\ket{vac}_1\otimes\ket{vac}_2$.
The measured observables are $\hat{X}_1\equiv(\hat{a}_1^\dagger+\hat{a}_1)/\sqrt{2}$ and $\hat{X}_2\equiv(\hat{a}_2^\dagger+\hat{a}_2)/\sqrt{2}$, and the measurement is done by the homodyne measurement with an infinitely squeezed reference light beam (i.e., the noise is reduced to zero). 
When quantum entanglement is not present, namely $\gamma=0$, this quantum structure faithfully represent the classical game, and the final measurement provides the original classical results: $p_1\equiv\braket{\psi_f|\hat{X}_1|\psi_f}=x_1$ and $p_2\equiv \braket{\psi_f|\hat{X}_2|\psi_f}=x_2$.
 Otherwise, namely when quantum entanglement is present, the prices charged by the two firms are determined by

\begin{eqnarray}
	p_1 &=& x_1\cosh \gamma +x_2\sinh \gamma,\\
	p_2 &=& x_2\cosh \gamma +x_1\sinh \gamma.
\end{eqnarray}
 Note that the classical model can be recovered by choosing $\gamma$ to be zero, since the two firms can directly decide their quantities. On the other hand, both $p_1$ and $p_2$ are determined by $x_1$ and $x_2$ when $\gamma\neq 0$. It leads to the correlation between the firms.

Similar to the classical version, in equilibria both firm must choose $\hat{x}\equiv\hat{p}e^{-\gamma}$ so that their prices are equal to the CE price (under each of the two rationing rules).
Obviously, under the efficient rationing rule, where the residual demand is independent from the opponent's price, the existence condition is the same as in the classical model.
In the following we consider the existence of the equilibrium in our quantum extension under the proportional rationing.
That is, we check whether a firm can be better off by changing her action provided that the opponent chooses $\hat{x}$.

Suppose that firm $i$ chooses $x_i$ and the opponent chooses $\hat{x}$.
Then, $i$'s price is given by, 
\begin{equation}
	\tilde{p}_i(x_i)\equiv x_i\cosh\gamma+\hat{x}\sinh\gamma.
\end{equation}
We assume $x_i\in [\hat{x},(a-\hat{x}\sinh\gamma)/\cosh\gamma)$, that is, $\tilde{p}_i(x_i)\in [\hat{p},a)$, otherwise $i$ will be worse off.
Firm $i$'s residual demand and profit are, respectively,
\begin{eqnarray}
	\tilde{R}^Q_i(x_i) &\equiv& D(\tilde{p}_i(x_i))\cdot g(x_i), \\
	\tilde{u}_i^Q(x_i) &\equiv& \pi(\tilde{p}_i(x_i))\cdot g(x_i), 
\end{eqnarray}
where
\begin{eqnarray}
	g(x_i) &\equiv& \left(1-\frac{k}{a-\hat{x}\cosh\gamma-x_i\sinh \gamma}\right), \\
	\pi(p) &\equiv& pD(p)=p(a-p).
\end{eqnarray}
Note that $g',g'',\pi''<0$, and that $\pi'(p)\le 0$ if and only if $p\ge a/2$.
In the classical model if $i$ raises her price, then her share will be $1-k/D(\hat{p})$ (independent from her price).
On the other hand, in the quantum model if $i$ raises her price, then the opponent's price also rises due to the entanglement.
And hence, her share will become $g(x_i)<1-k/D(\hat{p})$ if $i$ chooses $x_i>\hat{x}$.
Namely, it implies that her gain from raising price is always smaller than that in the classical model.
Therefore, the existence condition in the quantum model is more mild than that in the classical model.
In fact, as we will show below, given any $\gamma\ge 0$, $[\tilde{u}^Q_i(\hat{x})]'\le 0$ if and only if
\begin{equation}\label{Upper bound}
	k\le k(\gamma) \equiv \frac{ae^\gamma}{2(\cosh \gamma+e^\gamma)}.
\end{equation}
It should be worth noticing that since $\tilde{R}^Q_i(x_i)$ is not concave, the first order condition Eq. (\ref{Upper bound}) is not sufficient to guarantee the existence of the equilibrium.
However, we can show $[\tilde{u}^Q_i(x_i)]'\le 0$ for all $x_i$ if and only if $k\le k(\gamma)$.
Consequently, we obtain the following proposition.
\begin{pro}
For any $\gamma\ge 0$, $(\hat{x},\hat{x})$ is the unique quantum Nash equilibrium if and only if $k\le k(\gamma)$.
\end{pro}
\begin{pf}
The proof is consisted of the following four steps.

\begin{enumerate}
	\item Each firm will be worse off by deviating to $x_i$ such that $\tilde{p}_i(x_i)\notin [\hat{p},a)$.

On the one hand, even if $i$ selects $x_i$ such that $\tilde{p}_i(x_i)<\hat{p}$, she cannot increase her sales.
On the other hand, choosing $x_i$ such $\tilde{p}_i(x_i)\ge a$ implies her residual demand being zero.

	\item $[\tilde{u}^Q_i(x_i)]'\le 0$ for all $x_i$ such that $\hat{p}_i(x_i)\in[a/2,a)$.

Take any $x_i$ such that $\hat{p}_i(x_i)\in[a/2,a)$. Then,
\begin{equation}
	[\tilde{u}^Q_i(x_i)]'=\pi(\tilde{p}_i(x_i))g'(x_i)+\cosh \gamma \pi'(\tilde{p}_i(x_i))g(x_i)\le 0,
\end{equation}
since both terms in the RHS are nonpositive.

	\item $[\tilde{u}^Q_i(x_i)]''\le 0$ for all $x_i$ such that $\hat{p}_i(x_i)\in[\hat{p},a/2)$.

Take any $x_i$ such that $\hat{p}_i(x_i)\in[\hat{p},a/2)$. Then,
\begin{equation}
	[\tilde{u}^Q_i(x_i)]''=\pi(\tilde{p}_i(x_i))g''(x_i)+2\cosh \gamma \pi'(\tilde{p}_i(x_i))g'(x_i)+(\cosh\gamma)^2\pi''(\tilde{p}_i(x_i))g(x_i)\le 0,
\end{equation}
since each term in the RHS is nonpositive.

\item $[\tilde{u}^Q_i(\hat{x})]'\le 0$ if and only if $k \le k(\gamma)$.
 
Note that 
\begin{eqnarray}
	& & \tilde{p}_i(\hat{x})=\hat{p}=a-2k, \quad \tilde{p}'_i(\hat{x}) =\cosh\gamma, \quad \tilde{R}^Q_i(\hat{x})=k, \\
	& & [\tilde{R}^Q_i(\hat{x})]'=2k\left(-\frac{\sinh\gamma}{4k}\right)-\frac{1}{2}\cosh\gamma =-\frac{e^\gamma}{2}.
\end{eqnarray}
Since $u_i^Q(x_i)=\tilde{p}_i(x_i)\cdot \tilde{R}_i^Q(x_i)$, thus
\begin{eqnarray}
	[\tilde{u}^Q_i(\hat{x})]' & = & k\cosh\gamma-(a-2k)\frac{e^\gamma}{2}, \\
	& = & k(\cosh\gamma + e^\gamma) -\frac{a e^\gamma}{2},
\end{eqnarray}
which is nonpositive if and only if $k \le k(\gamma)$. $\Box$
\end{enumerate}
\end{pf}

\begin{figure}[ht]
   \begin{center}
     \input{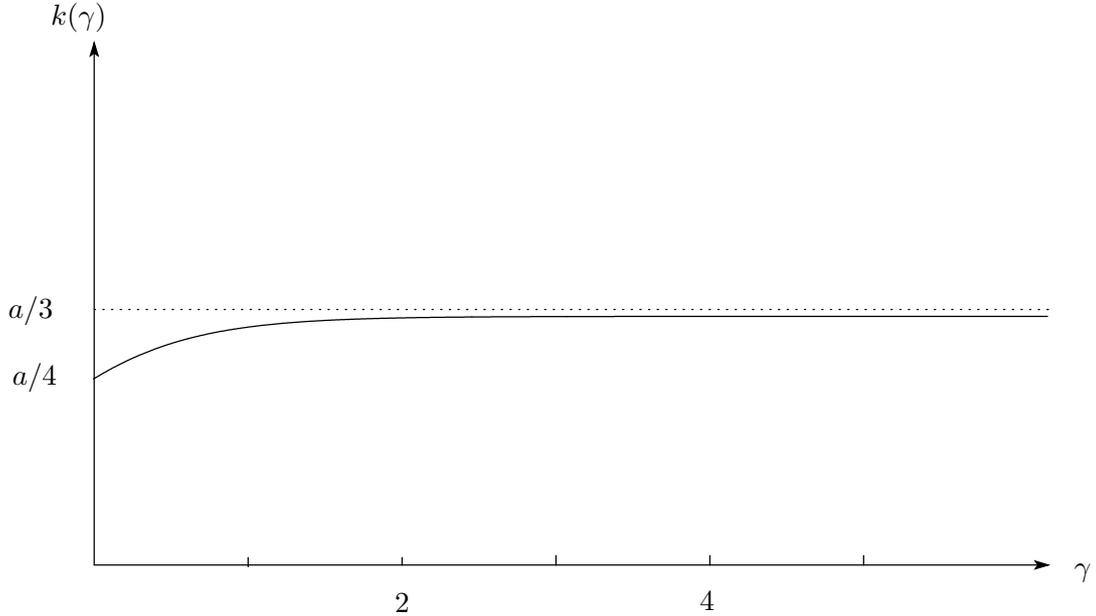}
     \caption{\label{fig} The upper bound of the capacity $k(\gamma)$ which ensures the existence of the quantum equilibrium as a function of the entanglement parameter $\gamma$.
     }
   \end{center}
\end{figure}

This proposition implies that the upper bound $k(\gamma)$ which ensures the existence of the equilibrium monotonically increases and converges to $a/3$ as $\gamma$ goes to $\infty$ (See Fig. \ref{fig}).
Thus, the existence problem is fully resolved in the maximum entanglement state.

\section*{Acknowledgements}
We all acknowledge financial support from Grant-in-Aid for Creative Scientific Research no 19GS0101 of the Japan Society for the Promotion of Science. 
TS was also supported by Grant-in-Aid for Young Scientists (Start-up) no 20830095 of the Japan Society for the Promotion of Science.

\section*{Appendix}
Here we show the uniqueness of equilibria.
\begin{pro}
In the classical game where the residual demand is determined by either the proportional rationing rule or the efficient rationing rule, if $(p_1,p_2)$ is a Nash equilibrium, then $p_1=p_2=\hat{p}$.
\end{pro}
\begin{pf}
Suppose that $(p_1,p_2)$ is a Nash equilibrium.
Obviously, we can assume $p_1,p_2\in(0,a)$, otherwise either firm obtains zero profit.
First, we show that $p_1=p_2$. Suppose to the contrary that $p_1\neq p_2$. 
We can assume WLOG $p_1>p_2$. 
On the one hand, if $R_1(p_1,p_2)=0$, then Firm 1 can be better off by changing her price to $p_2-\varepsilon$. 
On the other hand, if $R_1(p_1,p_2)>0$, then Firm 2 can be better off by raising her price slightly. 
Thus, both firms charge the same price $p^*$. 

Next, we show that $p^*=\hat{p}$. 
Suppose, to the contrary, that $p^*\neq \hat{p}$. 
On the one hand, if $p^*>\hat{p}$, then each firm can be better off by undercutting. 
On the other hand, if $p^*<\hat{p}$, then each firm can be better off by raising her price slightly. 
$\Box$
\end{pf}

Similar assertion is also valid in the quantum model, that is, if $(x_1,x_2)$ is a quantum Nash equilibrium, then $x_1=x_2=\hat{x}$.

\end{document}